\documentclass{article}

    \usepackage[preprint]{neurips_2023}

\usepackage[utf8]{inputenc} 
\usepackage[T1]{fontenc}    
\usepackage{hyperref}       
\usepackage{url}            
\usepackage{booktabs}       
\usepackage{amsfonts}       
\usepackage{nicefrac}       
\usepackage{microtype}      
\usepackage{xcolor}         

\usepackage{bm} 
\usepackage{graphicx} 
\usepackage{svg} 
\usepackage{subfig} 
\usepackage{xfrac} 

\title{From Data-Fitting to Discovery: Interpreting the Neural Dynamics of Motor Control through Reinforcement Learning}

\author{%
  Eugene R. Rush \\
  Department of Mechanical Engineering\\
  University of Colorado Boulder\\
  Boulder, CO 80309 \\
  \texttt{eugene.rush@colorado.edu} \\
  \And
  Kaushik Jayaram \\
  Department of Mechanical Engineering\\
  University of Colorado Boulder\\
  Boulder, CO 80309 \\
  \texttt{kaushik.jayaram@colorado.edu} \\
  \And
  J. Sean Humbert \\
  Department of Mechanical Engineering\\
  University of Colorado Boulder\\
  Boulder, CO 80309 \\
  \texttt{sean.humbert@colorado.edu} \\
}

\begin{document}

\maketitle

\begin{abstract}




In motor neuroscience, artificial recurrent neural networks models often complement animal studies. However, most modeling efforts are limited to data-fitting, and the few that examine virtual embodied agents in a reinforcement learning context, do not draw direct comparisons to their biological counterparts. Our study addressing this gap, by uncovering structured neural activity of a virtual robot performing legged locomotion that directly support experimental findings of primate walking and cycling. We find that embodied agents trained to walk exhibit smooth dynamics that avoid tangling -- or opposing neural trajectories in neighboring neural space -- a core principle in computational neuroscience. Specifically, across a wide suite of gaits, the agent displays neural trajectories in the recurrent layers are less tangled than those in the input-driven actuation layers. To better interpret the neural separation of these elliptical-shaped trajectories, we identify speed axes that maximizes variance of mean activity across different forward, lateral, and rotational speed conditions.

\end{abstract}

\section{Introduction}

Recurrent neural networks (RNNs) have become a powerful tool for computational neuroscientists in the past decade, primarily in modeling experimental neural data, and more recently as a means of constructing in-silico models that compliment findings from task-oriented animal research. A `computation through dynamics' neuroscience paradigm has emerged \cite{sussillo_opening_2013,saxena_towards_2019,vyas_computation_2020}, which views high-dimensional neural populations within biological and artificial RNNs as dynamical systems, $ \bm{dx} / \bm{dt} = f(\bm{x}(t), \bm{u}(t)) $, mapping external input $\bm{u}$, to neural states $\bm{x}$, via nonlinear recurrent connectivity $f$. Because artificial RNN models have full observability, they are being utilized more frequently to strengthen research findings within the neuroscience community \cite{sussillo_neural_2015}. 

However, advances in reinforcement learning (RL) have not yet been directly applied within the field of computational neuroscience to corroborate or extend experimental findings. This study aims to exemplify that deep RL, with its rich, closed-loop agent-environment interaction, can be a valuable tool for exploring computational principles found in embodied motor control systems.


Motor neuroscience researchers have found geometric neural properties during a repetitive cycling task that are conserved across primates species. Specifically, neural trajectories of the primate motor cortex do not cross over one another or tangle, while muscle-like electromyography (EMG) readings often do \cite{russo_motor_2018,saxena_motor_2022} during forward and backward cycling. Trajectory tangling is formally defined as $Q(t) = \max\limits_{t'} \Vert\dot{\bm{x}_{t}} - \dot{\bm{x}}_{t'} \Vert^2 / (\Vert {\bm{x}_{t} - \bm{x}_{t'} \Vert}^2 + \varepsilon ) $, and is a measure of how opposed trajectory velocities are relative to trajectory position. It is thought that highly-recurrent systems, such as motor cortices, exhibits low trajectory tangling as a means of avoiding potential instabilities, whereas systems driven by external signals such as muscles, may be highly tangled \cite{russo_motor_2018}. An intuitive way of understanding this is to consider that slight differences in muscle activation, can result in very different task behaviors, such as forward and backward cycling. Researchers extended this finding from simple opposing motion to speed modulation, showing biological and artificial RNN solutions reflect the principle that smooth well-behaved dynamics require low trajectory tangling \cite{saxena_motor_2022}.

In this paper, we demonstrate that these geometric neural properties are also conserved in agent-based reinforcement learning (RL) models. Unlike the standalone RNNs trained via supervised learning to fit EMG data \cite{russo_motor_2018}, RL models capture the rich, closed-loop interaction between agent and environment. Additionally, this approach empowers us to reach beyond neural recordings from animals, and embrace a completely new type of testbed for neuroscience, robotics, and machine learning research.

\subsection{Related Work}

\textbf{Task-oriented RNNs.} Many studies in recent years have focused on interpreting RNNs, such as those trained to perform text classification \cite{aitken_geometry_2022} and sentiment analysis \cite{maheswaranathan_how_2020,maheswaranathan_reverse_2019}, transitive inference \cite{kay_neural_2022}, pose estimation \cite{cueva_emergence_2018,cueva_emergence_2020,cueva_recurrent_2021}, memory \cite{cueva_low-dimensional_2020}, and other cognitive tasks \cite{yang_task_2019}. These models are powerful to neuroscientists and machine learning researchers alike, since they are transparent and allow full access to the internal state of the system. However, these works mainly train RNNs on open-loop cognitive tasks.

\textbf{Interpreting embodied behavior.} In general, there have been few efforts thus far to extend this work towards closed-loop control tasks, despite RL being a natural solution for connecting sensorimotor processing with goal-directed, embodied behavior. One exception is \cite{merel_deep_2020}, in which researchers developed a virtual rodent model as a testbed for studying neural activity across various high-level behaviors. Another is an in-silico study \cite{singh_emergent_2021}, which examined the population-level dynamics of a virtual insect navigating to the source of an an odor plume. The analyses in these studies reveal coordinated neural activity patterns, however neither make direct connections to motor neuroscience hypotheses: the former \cite{merel_deep_2020} focusing chiefly on features of multi-task neural behavior, and the latter \cite{singh_emergent_2021} focusing on how neural activity relates to spatial localization and navigation.

\textbf{High-performance robotics \& RL.} Conversely, in recent years the robotics community has demonstrated complex task-oriented behavior, such as legged locomotion \cite{rudin_learning_2022,lee_learning_2020,miki_learning_2022,feng_genloco_2023} and dexterous manipulation \cite{openai_learning_2019,handa_dextreme_2022}, but has a sparse literature on relating neural activity to embodied behavior. One major reason is that many high-performance RL systems utilize feed-forward networks and do not need RNNs, which are often less straightforward to train. These feedforward systems satisfy the Markov assumption -- that input observations completely describe the system state -- through frame stacking, where computer memory is used to store prior observations with current ones. Alternatively, for systems with periodic reward functions, a clock signal can be provided by the computer as temporal reference \cite{siekmann_sim--real_2021}. 

However, there are learned robotic controllers that have found major success in employing RNNs, such as bipedal locomotion \cite{siekmann_learning_2020}, quadrupedal locomotion over complex terrain \cite{rudin_learning_2022}, and dexterous manipulation \cite{openai_learning_2019,handa_dextreme_2022}. With the exception of \cite{siekmann_learning_2020,merel_deep_2020,singh_emergent_2021}, which found cyclic patterns of neural activity, there has been very little study of the connection between neural activity and embodied behavior of these in-silico RNN-based RL models. Overall, it is clear that advancements in deep RL have led to functional breakthroughs in robotics, but remains a fertile ground for computational neuroscience.

\section{Methods}

This work requires not only a high-throughput virtual experiment testbed, but also an analysis pipeline for studying the neural activity of embodied agents.

\subsection{Virtual Experiment Testbed}

\textbf{Agent \& Environment.} We utilize NVIDIA Isaac Gym \cite{makoviychuk_isaac_2021}, an end-to-end GPU-accelerated physics simulation environment, and a virtual model of the Anymal robot \cite{hutter_anymal_2016,rudin_learning_2022}, which is specified in IsaacGymEnvs\footnote{\href{https://github.com/NVIDIA-Omniverse/IsaacGymEnvs}{https://github.com/NVIDIA-Omniverse/IsaacGymEnvs}}.

The action space is 12-dimensional, with each of the four leg driven by three joint torque commands. The observation space is 176-dimensional, consisting of three translational body velocities $[u, v, w]$, three rotational body velocities $[p, q, r]$, three body orientation angles $[\theta, \phi, \psi]$, three planar body velocity commands $[u^{*}, v^{*}, r^{*}]$, 12-dimensional joint angle positions $\boldsymbol{\theta_{joints}}$, 12-dimensional joint angular velocity $\boldsymbol{\omega_{joint}}$, and 140-dimensional depth measurements $\boldsymbol{d}$.

To train the agent, the default reward function is a weighted sum of linear body velocity error $(u^{*}-u)^{2} + (v^{*}-v)^{2}$ and angular body velocity error $p^2 + q^2 + (r^{*}-r)^2$, along with a suite of knee collision, joint acceleration, change in torque, and foot airtime penalties. During training, agents are randomly assigned linear body velocity commands $[u^{*},v^{*}]$, as well as an angular velocity command $r^{*}$ that are modulated to regulate to a random goal heading. Additionally, these agents are trained using the AnymalTerrain curriculum learning, in which they gradually face more difficult various terrain obstacles as training progresses. To improve the robustness of the learned controller during training, agents are exposed to sensory noise, slight perturbations to gravity and friction, as well as random linear body velocity perturbations.

\begin{figure}[h]
    \centering
    \includegraphics[width=0.325\textwidth]{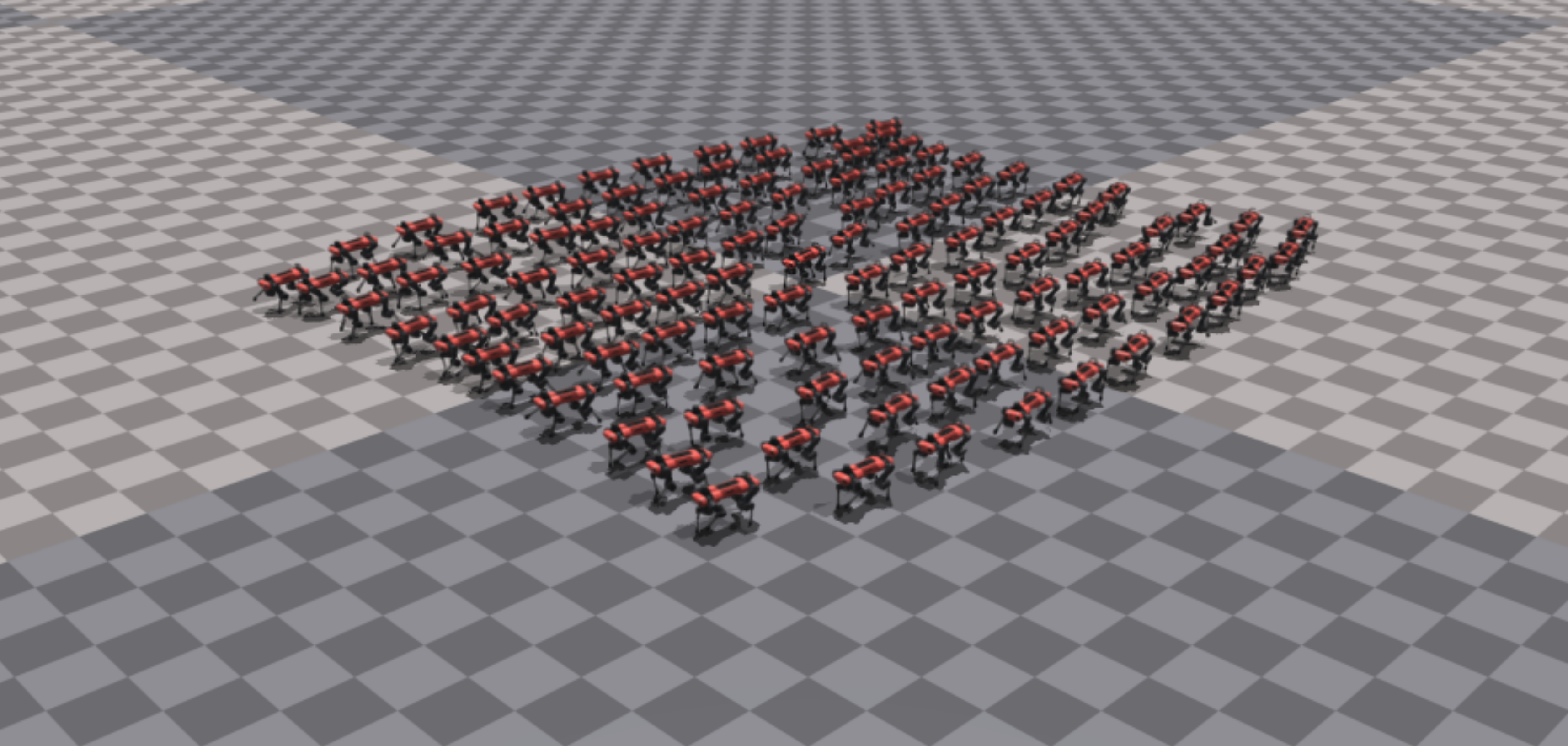}
    \includegraphics[width=0.325\textwidth]{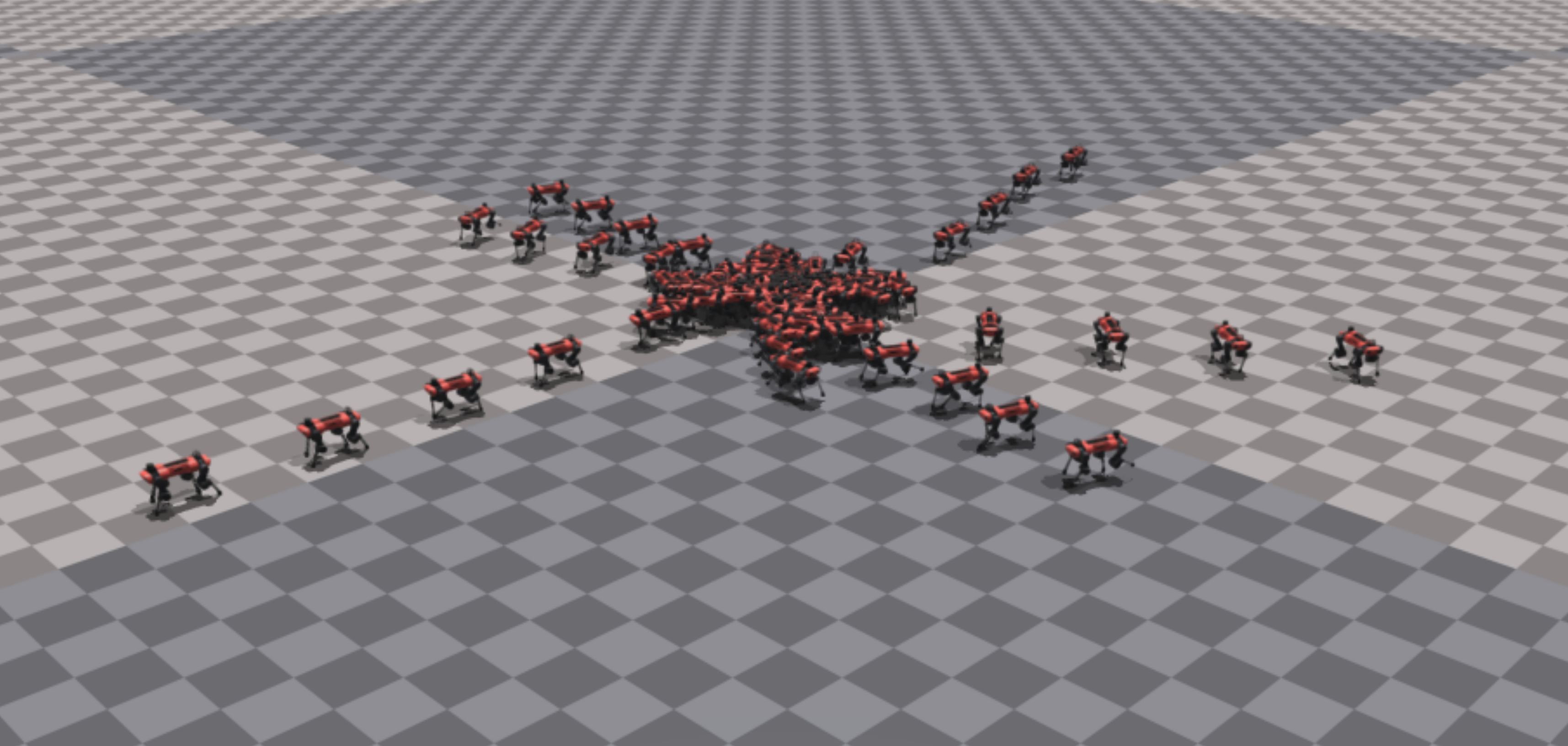}
    \includegraphics[width=0.325\textwidth]{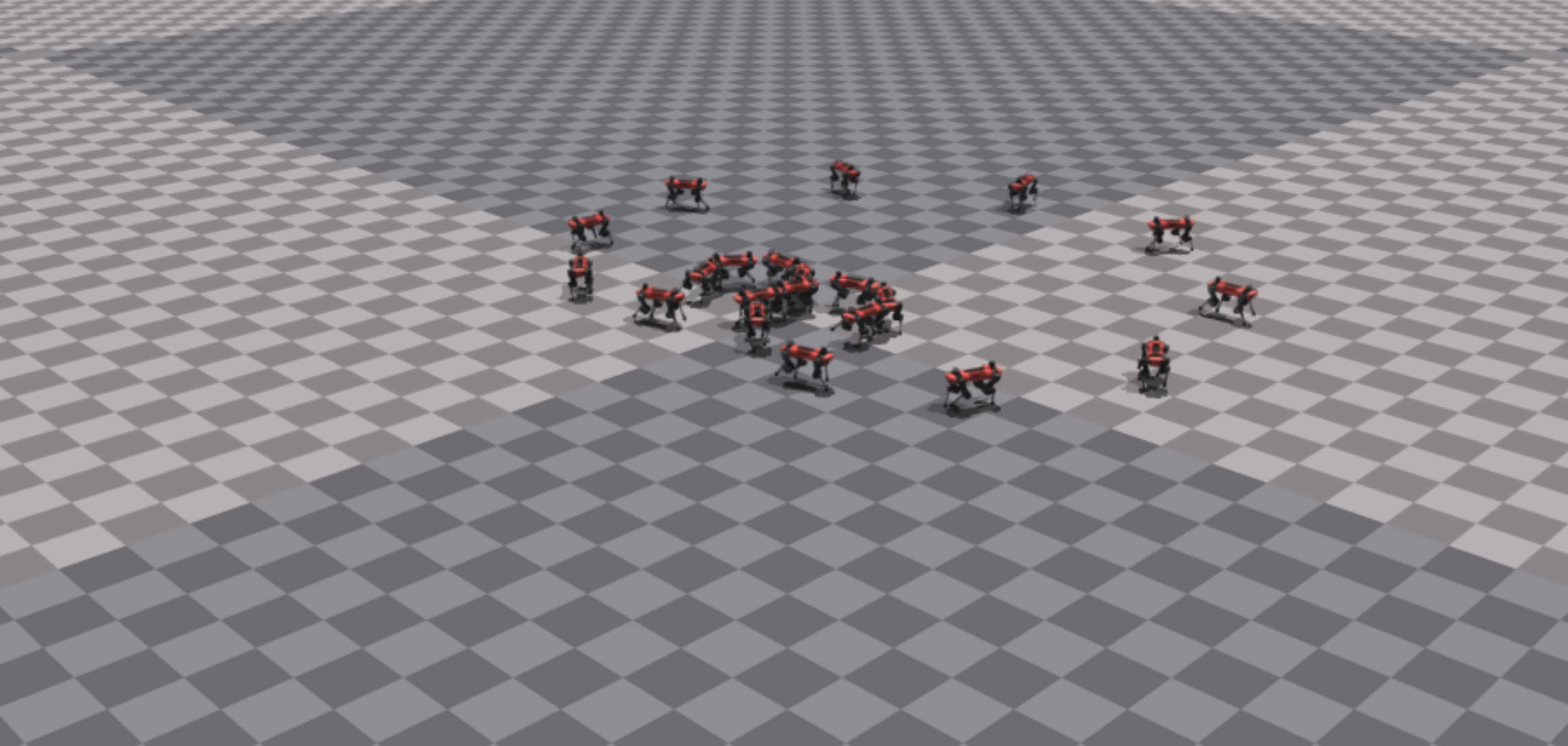}
    \caption{Rendering of three different experiments, where Anymal robots are walking each with a different forward, sideways, and rotational velocity commands in massively-parallelized NVIDIA Isaac Gym physics simulation engine.}
    \label{fig:agent_environment}
\end{figure}

After training, and during data collection, the agent is commanded with specified $[u^{*}, v^{*}, r^{*}]$, depending on the needs of the experiment. 
Some example velocity commands result in the physical configurations shown in Figure \ref{fig:agent_environment}. During data collection, noise parameters remain the same as during training, but perturbations are removed, except for during the perturbation experiment.

\textbf{Model Architecture.} The agent is trained using rl\_games\footnote{\href{https://github.com/Denys88/rl_games}{https://github.com/Denys88/rl\_games}}, a high performance RL libary that implements Asymmetric Actor Critic (A2C) \cite{pinto_asymmetric_2017}, which is a variant of Proximal Policy Optimization (PPO) \cite{schulman_proximal_2017}. We utilize a implementation that integrates Long-Short Term Memory (LSTM) networks into the actor and critic networks. Both actor and critic networks pass the 176-dimensional observation vector through a multi-layer perceptron with two dense layers of size 512 and 256, a single-layer LSTM network with 128 cells, and output a 12-dimensional action vector.

The neural activations of these cell states $[128 \times 1]$ is referred to as the RNN latent state or recurrent activity. The activations in the action vector $[12 \times 1]$ are referred to as the actuation activity or actuation torque. These network populations are the primary focus of this work, and are compared to the motor cortex and muscle-like EMG recordings in primate studies \cite{russo_motor_2018,saxena_motor_2022}, respectively. Activity of the observation vector $[176 \times 1]$ and LSTM hidden state output vector $[128 \times 1]$ are also relevant, but the not the focus of this work, and are visualized in the Supplementary Materials. The critic network has identical structure, in addition to a dense layer that outputs a 1-dimensional value estimate. A diagram of the network layout is provided in the Supplementary Material.


The rl\_games implementation utilizes truncated backprogogation through time (BPTT) for training the RNNs. With a truncation length of four, this essentially unfolds the RNN into four layers, upon which gradients are propagated back through the network for each time step. Generally, BPTT truncation aids in training because it prevents long sequences from causing vanishing or exploding gradient issues.







\subsection{Analysis}

With the methodology described in the prior section, we now have agents that have been trained to walk with a wide variety of gaits, depending on the input body velocity commands. With NVIDIA Isaac Gym, we can take a single trained policy and can simulate hundreds to thousands of agents executing that policy simultaneously. In this section, we describe the analysis methodologies utilized to interpret the data recorded from the experimental trials.

\textbf{Cycle alignment and averaging.} Through training, our deep RL agents learned how to walk at specified combinations of forward, lateral, and rotational body velocities. Since agents were relatively well-behaved and measured velocities were on average within 10\% of their reference velocities, we clustered trials based on input velocities.

Trials are aligned, so that they begin when the front right foot lifts off the ground, marking the beginning of the swing phase. The trials ends during after a full gait cycle, which is the last moment of the same leg's stance phase. This provides a consistent temporal reference point across different trials and conditions. Because of the stochastic nature of the agent's control policy, its gait cycle behavior vary slightly in duration, even with identical reference velocities. Typically, alignment can be achieved through non-uniform stretching in time, where data is aligned according to the respective gait positions. However, with the ease of virtual data collection in a massively parallelized simulation environment, we simplified data alignment by simply computing the most common gait cycle period for each condition, and averaging data across those trials according to time step.

Data was recorded from the system at a simulation frame rate of $50\text{Hz}$ or every $20\text{ms}$. Because gait cycles varied between $1-2.5\text{Hz}$, each cyclic trajectory consists of $20-50$ data points, depending on the walking speed. To improve the interpretability of our three-dimensional visualizations, we utilized cubic-spline interpolation with $1\text{ms}$ resolution on the neural data in the first three PCs, as well as tangling values.

\textbf{Computing tangling.} Trajectory tangling, $Q(t)$, is a measurement that helps determine the complexity of neural or muscle state dynamics over time. It is computed as $Q(t) = \max\limits_{t^{\prime}} \Vert\dot{\bm{x}}_{t} - \dot{\bm{x}}_{t^{\prime}} \Vert^2 / (\Vert {\bm{x}_{t} - \bm{x}_{t^{\prime}} \Vert}^2 + \varepsilon ) $. Here, $\bm{x}(t)$ represents the neural or muscle state at a specific time $t$, and $\bm{x}(t)$ is its temporal derivative. The Euclidean norm is denoted as $\|\cdot\|$, and $\varepsilon$ is a constant that prevents division by zero. The derivative of the state, $\dot{\bm{x}}(t)$, is simply computed as the difference between the state at times $t$ and $t-\Delta t$, divided by $\Delta t$. The constant $\varepsilon$ is set to $0.1$ times the variance of $\bm{x}$, which prevents the denominator from going below $0.1$ times the average squared distance from zero.

Before computing tangling, we mean-center and scale the average cyclic trajectories, so that each neural dimension has zero mean and unit variance, with respect to the entire dataset. Note that we do not compute tangling after performing principal component analysis (PCA) on the data, though the results should be similar. Since tangling is computed relative to other measurement samples in a dataset, we can  compute tangling to study behavior within a single trajectory, or compute it relative to other trajectories. In this paper, we do both depending on whether the experiment focuses on a single or multiple conditions.

\textbf{Principal component analysis by speed and across speeds.} Data is preprocessed via mean-centering and scaling, so that each neural dimension has zero mean and unit variance, before conducting principal component analysis (PCA). PCA inherently captures maximum variance in as few dimensions as possible, computing and applying independent transformations for neural and actuation network populations. In addition to conducting PCA on single-speed trials, we also use it to compute the `global subspace,' which explains the response variance across a range of velocity conditions. To do this, we simply aggregated the trial data from different conditions into a single matrix and utilize the same process as in single-speed PCA.



\textbf{Computing speed-varying axis.} This speed axis is orthogonal to PCs 1 and 2 and maximizes the variance of the average neural activity across speeds, which is denoted by $\bm{X}_{s} = \frac{1}{T} \sum_{t}^{T_{s}} x_{s}(t)$, where $x_{s}(t)$ is neural data for speed $s$ at time $t$, projected into the remaining across-speed PCs. In general terms, we maximize the cost function $\|\bm{X}\bm{d}\|$, where $\bm{X}$ $[S \times D]$ is a known matrix consisting of the $S$ speed-specific, cycle-averaged RNN activations transformed into PC $3$ through PC $(D+2)$ space, and $\bm{d}$ $[D \times 1]$ is an unknown vector of coefficients with constraint $\| \bm{d} \| = 1$. In our case, $S=7$ and $D=10$, and this procedure finds $\bm{d}$ $[10 \times 1]$ that represent the linear combination of PC $3$ through PC $12$ that maximizes the mean RNN activations across the 7 speed conditions. This leads to a modified transformation that yields the maximum separation of neural trajectories across speed conditions.

\section{Results}

First, we present the neural activity and respective intra-cycle tangling for a single condition: forward walking at $2\sfrac{\text{m}}{s}$. We next  examine inter-condition tangling across opposing motions, as it relates to \cite{russo_motor_2018}. To understand the neural activity and tangling during modulation of speed, and relate it back to prior work \cite{saxena_motor_2022}, we perform a third experiment across a range of speed conditions.



\begin{figure}[h]
    \centering
    \subfloat{
        \includegraphics[width=0.98\textwidth]{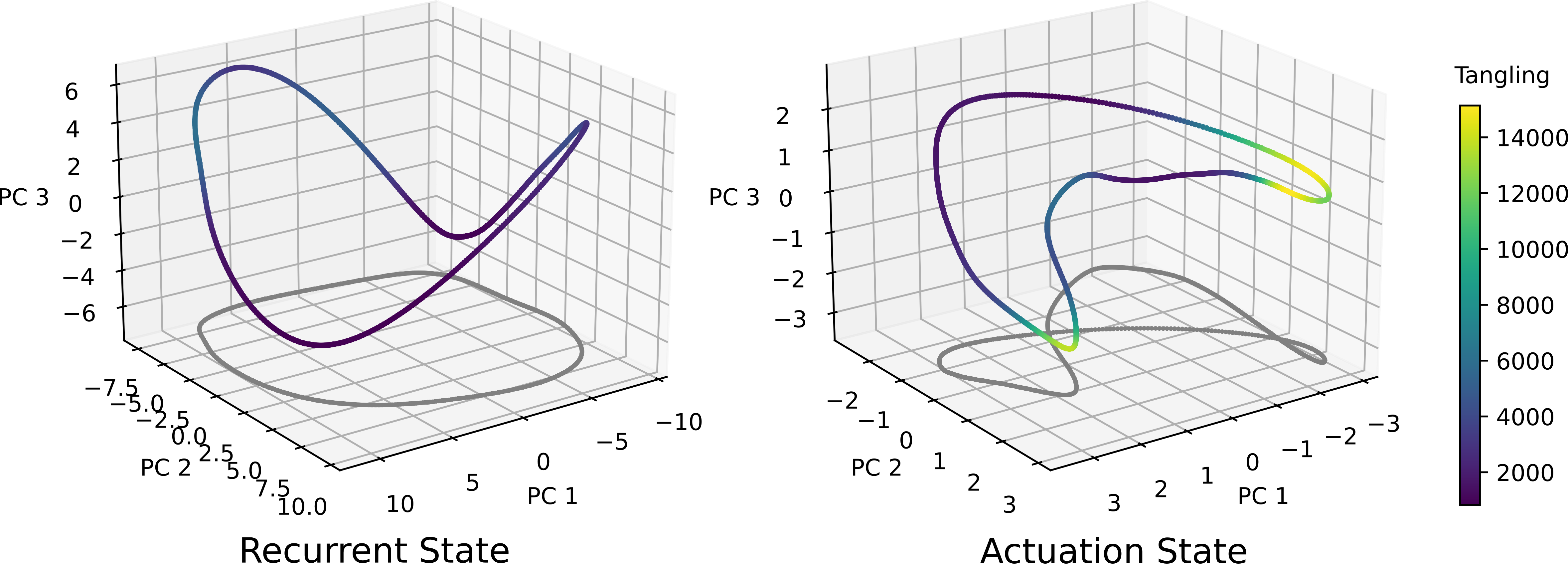}
    }
    \caption{Neural network activity during forward walking at $2\sfrac{m}{s}$. There are salient patterns that arise, (left) the latent recurrent state display a smooth, low-tangled trajectory with a nearly circular projection. Meanwhile, the actuation state (right) exhibits two sharp regions in its trajectory, causing high degrees of tangling, as well as a `figure eight'-shaped projection.}
    \label{fig:exp1_single_speed}
\end{figure}

\textbf{Recurrent activity exhibits lower intra-cycle tangling than actuation.} We study the recurrent state and actuation state of the agent during forward walking. When visualizing the recurrent state in its first three PCs, a salient elliptical-like pattern emerges in Figure \ref{fig:exp1_single_speed}. This aligns with the primate and artificial RNN behavior seen in \cite{russo_motor_2018}, which display smooth, nearly circular trajectories relative to the highly-tangled actuation pattern. Conversely, we see less intra-cycle separation in the actuation trajectory as it produces a `figure-eight' pattern when projected on its first two PCs. Additionally, it contains much sharper regions of its trajectory relative to the recurrent state, and experiences higher tangling as a result.

Circles are the least-tangled rhythmic trajectory, and it has been shown empirically that artificial RNNs trained to follow a low-tangling trajectory are more robust to noise \cite{russo_motor_2018}. High tangling was demonstrated by showing a particular class of Lissajous curves, $x=[\sin(t), \beta \cos(t), \cos(2t)]$, which forms a two-dimensional `figure eight'-shaped trajectory that intersects with itself when $\beta=0$. As $\beta$ is increased, the introduction of a third dimension increases separation increases between that cross over point, therefore reducing tangling. The projection in the first two dimensions, $[\sin(t), \beta \cos(t)]$, shifts from an ellipse when $\beta < 1$, to a circle when $\beta=1$.


\begin{figure}[h]
    \centering
    \subfloat{
        \includegraphics[width=0.98\textwidth]{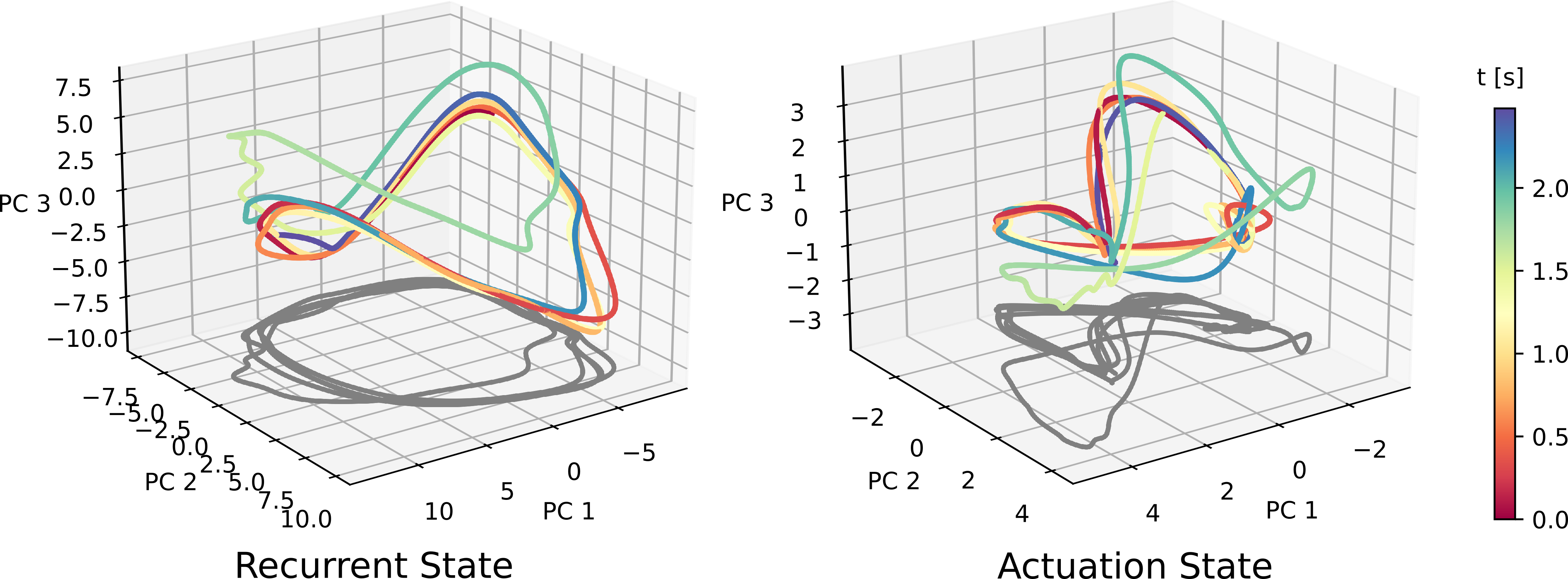}
    }
    \caption{Perturbation study of agent while walking forward at $2\sfrac{m}{s}$ reveals (left) the recurrent state moves in a direction orthogonal to its direction of motion and ultimately converging back to its nominal trajectory within nearly one cycle. This confirms that this structured trajectories reflect stable limit cycles. Specifically, at $t=1.5s$ the perturbation is applied for a single time step, after which the recurrent and actuation trajectories experience transient responses. This data is for a single agent, $N=1$, so there is no across-trial averaging.}
    \label{fig:exp1a_single_speed_perturb}
\end{figure}

We also conducted a variant of this first experiment, where we perturbed the agent with a random change in linear body velocity. After receiving this virtual `push', the agent reacts to regain balance, altering its recurrent and actuation states to compensate, as shown in Figure \ref{fig:exp1a_single_speed_perturb}. Within one to two gait cycles, the recurrent state converges back to its nominal cyclic trajectory. This confirms that the elliptical trajectory formed by the recurrent state in Figure \ref{fig:exp1_single_speed} represents a stable limit cycle. We explore in the next section, the input-dependent nature of these recurrent state limit cycles.

\textbf{Separation of recurrent activity during opposing motions.} This experiment consisted of comparing recurrent neural activity and actuation output during opposing locomotion behavior. We studied six conditions: walking forward and back $(u^{*} = \pm 2 \sfrac{\text{m}}{s})$, walking left and right $(v^{*} = \pm 2 \sfrac{\text{m}}{s})$, and turning left and right  while walking forward $(u^{*} = 2 \sfrac{\text{m}}{s},r^{*} \pm 0.25 \sfrac{\text{rad}}{s})$. For each condition, we simulated 100 agents, recorded data every $20\text{ms}$ for a total of $10\text{s}$, and averaged data according to cycle timing and condition.

\begin{figure}[h]
    \centering
    \subfloat{
        \includegraphics[width=0.80\textwidth]{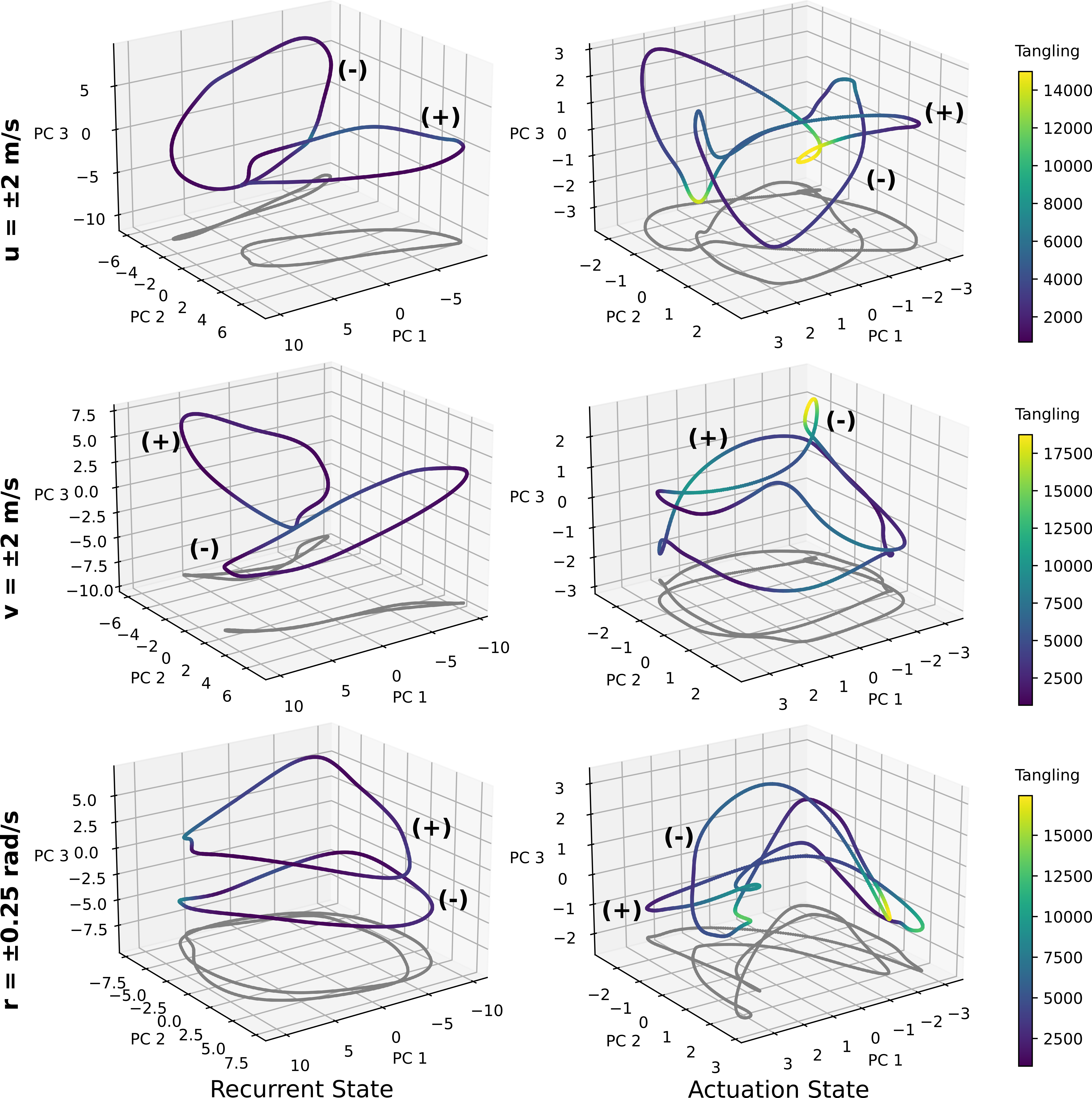}
    }
    \caption{Neural activity for body velocity commands of (top) walking forwards (+) and backwards (-), (middle) walking sideways left (+) and right (-), and (bottom) turning left (+) and right (-) during forward walking. Recurrent state  have low tangling and high separation in state space, whereas actuation states are highly tangled and appear to contain symmetries that cause trajectory overlap. Cyclic trajectories are grouped according to condition and intra-cycle time, averaged across all time and agents. For each of the seven conditions, $T=10\text{s}$, $N=100$.}
    \label{fig:exp2_opposingspeeds}
\end{figure}

The results shown in Figure \ref{fig:exp2_opposingspeeds} agree with the principle from \cite{russo_motor_2018}, that recurrent systems exhibit low tangling, while externally-driven systems can show higher degrees of trajectory tangling. The reasoning being that recurrent systems trained to be robust to noise, will become less tangled to avoid potential instabilities. In contrast, networks that do not rely on internal dynamics, such as sensory and actuation systems, can exhibit tangled trajectories when driven by different external signals. When comparing opposing motions in Figure \ref{fig:exp2_opposingspeeds}, joint torque trajectories display some overlapping symmetry, leading to low separation in state space. Additionally, rapid changes in actuation direction also contribute to this high-tangling characterization.

In primates, it was observed that muscle-like EMG signals were counter-rotating for forwards and backwards cycling, but co-rotating in the motor cortex \cite{russo_motor_2018}. However, here we see a significant deviation from biology, where in-silico trajectories of both the actuation system and recurrent systems are counter-rotating. Counter-rotation has the potential to lead to higher tangling and greater changes of instability in recurrent systems. Despite this, the RNN trajectories still displays relatively low tangling because there is significant translational separation in the neural state space.

\begin{figure}[h]
    \centering
        \includegraphics[width=0.99\textwidth]{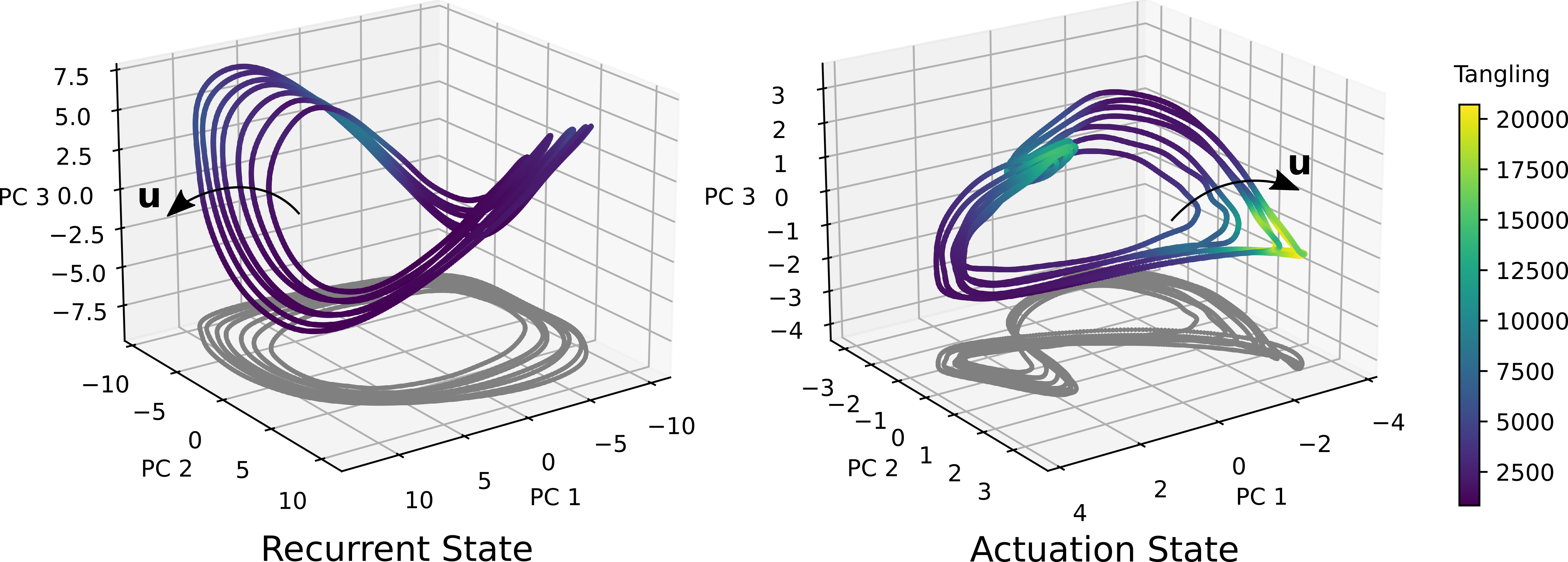}
    \caption{Neural data for a suite of forward speed commands, ranging from $0.8\sfrac{\text{m}}{s}$ to $2.0\sfrac{\text{m}}{s}$. Across-speed PCA was conducted on aggregated data from all seven conditions. For the recurrent state, the elliptical shape is retained as seen in Figure \ref{fig:exp1_single_speed}, as is the `figure-eight'-shaped projection in the actuation trajectory. At faster walking speeds, both the recurrent and actuation state seem to scale or `stretching', generating a separation of trajectories. Note that arrows indicate the direction of increasing forward velocity command.} 
    \label{fig:exp3_modulatespeed}
\end{figure}

\textbf{Emergence of speed-dependent limit cycles.} In this third experiment, we expand from simple inversion of velocity commands, to a modulation between slower and faster velocity commands. We compare neural activity and motor actuation across seven cases, $u^{*} = [0.8 , 1.0, 1.2, 1.4, 1.6, 1.8, 2.0]$ $\sfrac{\text{m}}{s}$. Figure \ref{fig:exp3_modulatespeed} reveals the recurrent state limit cycles maintain their shape as speed is modulated, but appear scaled or `stretched' as speed increases. The actuation state also shows increased scaling at faster speeds. This leads to separation between trajectories of different speeds, with no noticeable areas of overlap when projected into the top three principal component directions. We find similar patterns when varying $v$ and $r$ as well, and share those results in the the Supplementary Materials.

We also collect and study data during a similar experiment, but on an agent trained with actuation feedback. This is common in the some RL implementations, where the action vector, in this case 12 joint torque actuator commands, is concatenated with the observation vector and input to the neural network. When this is done, we observe greater tangling, as shown in the Supplemental Materials.

This separation of trajectories in state space is the expected solution for adjusting speed \cite{maheswaranathan_universality_2019,sussillo_opening_2013}, since additive inputs to RNNs cannot scale their respective flow-field magnitude. Therefore, it is required to move to a different part of the state space for the RNN to change trajectory speed \cite{remington_flexible_2018}. To note, the gait cycle frequency of our agent is roughly $1.8\text{Hz}$ when walking forward at $0.8\sfrac{\text{m}}{s}$, and $2.3\text{Hz}$ when walking at $2.0\sfrac{\text{m}}{s}$, which is roughly a 30\% increase in mean trajectory speed. This indirect mapping from gait frequency to walking speed should be considered, when comparing to the behaviorally more straightforward cycling task.

This leads to the second reason for trajectory changes as locomotion varies. Motor actuation activity changes magnitude, shape, and temporal pattern based on the velocity command, it doesn't simply speed up at faster locomotion speeds. Additionally, as seen in Figure \ref{fig:exp2_opposingspeeds}, motor activity has significantly different trajectory patterns depending on the body linear velocity direction $(u, v)$ and angular body velocity $r$.

While \cite{saxena_motor_2022} found this trajectory separation appeared as a `stack' of elliptical limit cycles, our results do not appear to exhibit the same characteristic when projected onto the top three PCs. The following section addresses this challenge, by transforming the data from this same experiment in order to better visualize this speed-dependent separation.

\begin{figure}[h]
    \centering
        \includegraphics[width=0.99\textwidth]{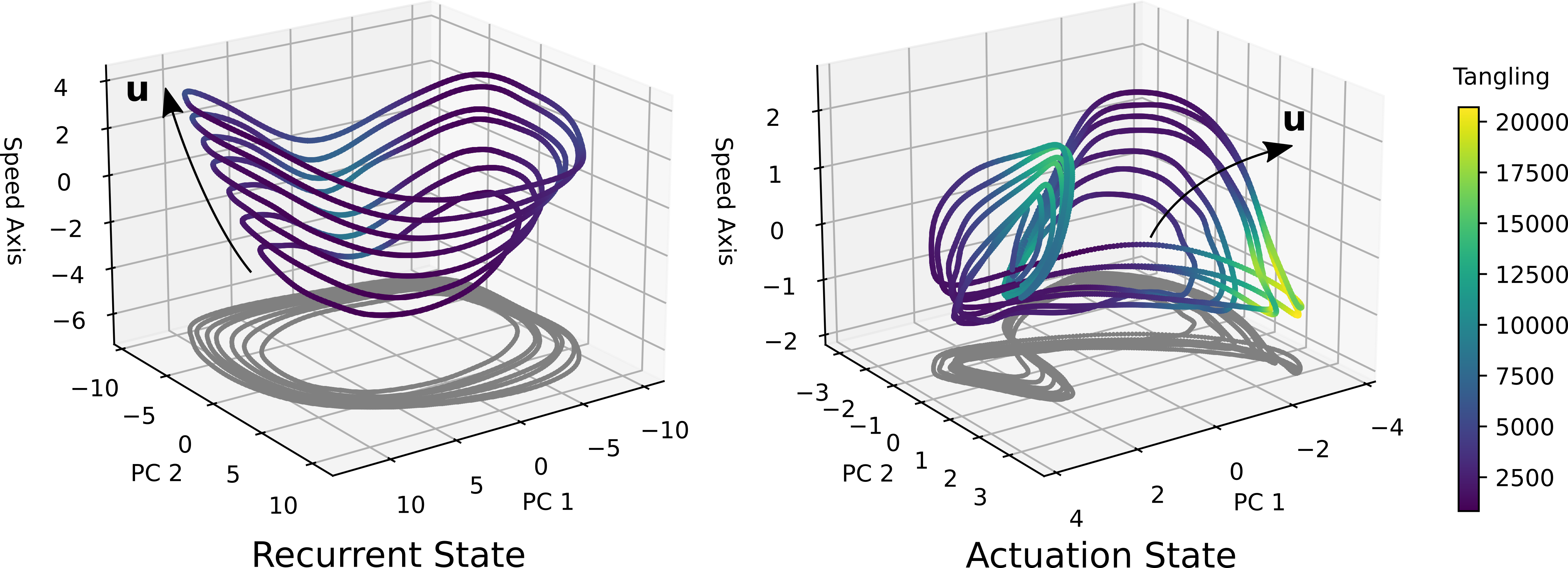}
    \caption{Neural data for various speed conditions projected along PC1, PC2, and Speed Axis. There is greater separation between the trajectories in the Speed Axis, as compared with PC3 shown in Figure \ref{fig:exp3_modulatespeed}. This is because the Speed Axis was optimized as the linear combinations of PC3-12 that yields the greatest variance between the mean neural activity of different speed-dependent trajectories. In many cases, the Speed Axis predominantly consists of PC3 and PC4, which contains variance that is not captured in Figure \ref{fig:exp3_modulatespeed}.} 
    \label{fig:exp3_modulatespeed_speed_axis}
\end{figure}

\textbf{Elliptical limit cycles are stacked along speed-dependent axis.} Since we are restricted to three-dimensional spatial visualizations, we borrowed inspiration from \cite{saxena_motor_2022}, and replaced PC3 with a speed axis, as shown in Figure \ref{fig:exp3_modulatespeed_speed_axis}. Described in the Methods section, we find the direction within the neural state space that maximizes the variance of the speed-averaged neural activity. We constrain the speed axis so it is orthogonal to PCs 1 and 2, while also restricting it to be a linear combinations of PCs 3 - 12. Interestingly, the fourth and fifth PC directions are in some case large contributors to the speed axis, meaning that we are now capturing neural variance that could not be visualized in Figure \ref{fig:exp3_modulatespeed}. 

Additional data on the speed axis characterizations is included in the Supplementary Materials. We also obtain similar results with speed-modulated $v$ and $r$ experiments, which are shared in the Supplementary Materials. Note that the speed axis is of less benefit to the actuation trajectories, which are still visibly tangled.

In conclusion, this analysis elucidates that recurrent dynamics display a speed-dependent stacked-elliptical solution, which results in low intra-cycle and across-speed tangling. Additionally, in the Supplementary Materials, we share a comparison of results across agents trained independently. During initialization the neural network parameters are randomized, resulting in different absolute neural activation patterns, but structured neural activity that is nearly identical to that shown for this agent.

\section{Discussion}

\textbf{Lower tangling in RNNs supported by embodied RL agents.} In this paper, we found neural network activity of robotic agents trained to walk has cyclic structure that resembles the neural activity in monkeys during cyclic activities, including cycling \cite{russo_motor_2018,saxena_motor_2022} and walk \cite{d_foster_freely-moving_2014}. Specifically, we see lower levels of trajectory tangling in RNN latent states than the rest of the network. The latent RNN neural activity is input velocity-dependent, effectively creating greater separation across trajectories. This result is significant, because it represents the first time this phenomenon has been identified and characterized in a embodied RL context. Previous in-silico studies are limited to artificial RNNs trained to output muscle-like EMG data from monkey trials. This study closes the loop, capturing the interaction between the agent and its environment.

\textbf{Tangling extends to tasks with higher-dimensional velocity commands.} In this study, we explore the locomotion task, which unlike cycling, is not one dimensional. For in-plane motion, the learned legged locomotion controller is governed by a three-dimensional input: forward, lateral, and rotational velocity commands. This goes beyond the primate study on forward walking \cite{d_foster_freely-moving_2014}, which found cyclic trajectories for slow and fast walking had low degrees of tangling. This study corroborates that finding, and expands upon it by studying the relationship between neural activity while also modulating lateral and rotational velocity. This example is representative of how robotics and virtual embodied systems can be leveraged to support and even advance efforts in computational neuroscience. 

\textbf{Greater experimental freedom with virtual embodiment.} There are significantly more experimental capabilities, when moving from data-fitting efforts to embodied RL systems. In this work, we showed an example of a physical perturbation, and the observed transient effects on the neural dynamics. Additionally, with the modeling of the close-loop system, we can capture how perturbations affect physical behavior. This was chosen as the primary perturbation study, since it represents a type of experiment that is not possible with prior data-fitting RNN studies. In those prior works, such an effort would require the experimenter to conduct additional live animal trials and collect the necessary neural data to re-train the RNN. Here, with a virtual embodiment and environment, running a physical perturbation test is no different than a standard baseline experiment.

Additionally, we can also vary parameters of the agent, such as adding weight to the agent, changing joint torque dynamics, or damaging individual joints. Finally, the environment or task itself can be manipulated, for example modifying the ground terrain structure or surface friction. With orders-of-magnitude faster throughput than working with live animals, embodied RL systems offer significant accessibility to computational neuroscience researchers. 


\textbf{Bridging task-agnostic and task-specific paradigms.} This work aims to inspire deeper interpretation of learned motor controllers within the machine learning community, through the application of computational neuroscience methodologies. We are excited by the increasing agility of legged robots within the context of mixed and variable environments, and expect significant progress in understanding these relatively novel systems. In general, research that synthesizes advancements in machine learning and computational neuroscience have potential to bridge the gap between the relatively task-agnostic and task-specific paradigms. It is significant that we observed network activity patterns during locomotion in this work, that resembled that seen during cycling \cite{russo_motor_2018,saxena_motor_2022}, a substantially different task. This work aims to exemplify to the computational neuroscience community the exploratory power of leveraging deep RL and physics simulation engines in the study of embodied neural systems.

\clearpage

\section*{Acknowledgements} 

We thank Satpreet Singh and Christopher Cueva for early creative conversations, Christopher Bate for early software support, Christoffer Heckman and Alessandro Roncone for helpful feedback, Shreya Saxena for reviewing the abstract and providing constructive input, Hari Krishna Hari Prasad for discussions, and Angella Volchko for providing paper comments. We are grateful for the technical support from Denys Makoviichuk, the author of the open-source RL implementation rl\_games used in this paper, as well as the creators of NVIDIA Isaac Gym and IsaacGymEnvs.




\bibliographystyle{alpha} 
\bibliography{references}  

\clearpage

\end{document}